\documentclass{pasj00}

\def\commenta{$^*$}
\def\commentb{$^\dagger$}
\def\commentc{$^\ddagger$}
\def\commentd{$^\S$}

\newcounter{author}
\def\authorcount#1#2{\refstepcounter{author}\label{#1}
                     \altaffiltext{\ref{#1}}{#2}}

\begin{document}
\SetRunningHead{T. Kato et al.}{CC Sculptoris: }

\Received{201X/XX/XX}
\Accepted{201X/XX/XX}

\title{CC Sculptoris: Eclipsing SU UMa-Type Intermediate Polar}

\author{Taichi~\textsc{Kato},\altaffilmark{\ref{affil:Kyoto}*}
        Franz-Josef~\textsc{Hambsch},\altaffilmark{\ref{affil:GEOS}}$^,$\altaffilmark{\ref{affil:BAV}}$^,$\altaffilmark{\ref{affil:Hambsch}}
        Arto~\textsc{Oksanen},\altaffilmark{\ref{affil:Nyrola}}
        Peter~\textsc{Starr},\altaffilmark{\ref{affil:Starr}}
        Arne~\textsc{Henden},\altaffilmark{\ref{affil:AAVSO}}
}

\authorcount{affil:Kyoto}{
     Department of Astronomy, Kyoto University, Kyoto 606-8502}
\email{$^*$tkato@kusastro.kyoto-u.ac.jp}

\authorcount{affil:GEOS}{
     Groupe Europ\'een d'Observations Stellaires (GEOS),
     23 Parc de Levesville, 28300 Bailleau l'Ev\^eque, France}

\authorcount{affil:BAV}{
     Bundesdeutsche Arbeitsgemeinschaft f\"ur Ver\"anderliche Sterne
     (BAV), Munsterdamm 90, 12169 Berlin, Germany}

\authorcount{affil:Hambsch}{
     Vereniging Voor Sterrenkunde (VVS), Oude Bleken 12, 2400 Mol, Belgium}

\authorcount{affil:Nyrola}{
     Nyrola observatory, Jyvaskylan Sirius ry, Vertaalantie
     419, FI-40270 Palokka, Finland}

\authorcount{affil:Starr}{
     Warrumbungle Observatory, Tenby, 841 Timor Rd,
     Coonabarabran NSW 2357, Australia}

\authorcount{affil:AAVSO}{
     American Association of Variable Star Observers, 49 Bay State Rd.,
     Cambridge, MA 02138, USA}



\maketitle

\begin{abstract}
We observed the 2014 superoutburst of the SU UMa-type
intermediate polar CC Scl.  We detected superhumps
with a mean period of 0.05998(2)~d during the superoutburst
plateau and during three nights after the fading.
During the post-superoutburst stage after three nights,
a stable superhump period of 0.059523(6)~d was detected.
We found that this object is an eclipsing system
with an orbital period of 0.058567233(8)~d.
By assuming that the disk radius in the post-superoutburst
phase is similar to those in other SU UMa-type dwarf novae,
we obtained a mass ratio of $q$=0.072(3) from
the dynamical precession rate of the accretion disk.
The eclipse profile during outbursts can be modeled by
an inclination of \timeform{80.6D}$\pm$\timeform{0.5D}.
The 2014 superoutburst was preceded by a precursor
outburst and the overall appearance of the outburst
was similar to superoutbursts in ordinary SU UMa-type
dwarf novae.  We showed that the standard thermal-tidal
instability model can explain the outburst behavior in
this system and suggest that inner truncation of the disk
by magnetism of the white dwarf does not strongly affect
the behavior in the outer part of the disk.
\end{abstract}

\section{Introduction}

   Cataclysmic variables (CVs) are close binary systems
consisting of a white dwarf and a red (or brown) dwarf
transferring the gas via Roche overflow
[for a general review of CVs, see \citet{war95book};
\citet{hel01book}].  Dwarf novae (DNe) are a subclass
of CVs that show outbursts.  SU UMa-type dwarf novae
are a class of DNe that show long-lasting superoutbursts
in addition to short, normal outbursts.  During superoutbursts,
superhumps, which have a period a few percent longer than
the orbital period, are observed.
It is widely believed that outbursts in DNe are caused
by thermal instability of the accretion disk
and superoutbursts and superhumps are caused by
tidal instability arising from the 3:1 resonance
in the accretion disk with the orbiting secondary
[thermal-tidal instability (TTI) model;
\citep{osa89suuma}; see \citet{osa96review} for
a review].  Quite recently, this picture has become
even more firmly established by analyses of Kepler
observations (\cite{osa13v1504cygKepler};
\cite{osa13v344lyrv1504cyg}).

   Although the TTI model does
not explicitly consider the effect of magnetism
of the white dwarf, some CVs have magnetic
fields of the white dwarf strong enough to affect
the dynamics in the accretion disk.  If the magnetic
field is strong enough, the accretion disk cannot form
and the transferred matter directly accretes on
the magnetic poles of the white dwarf.  This condition
is usually met in polars (AM Her-type objects) in which
the strong magnetic field synchronizes the rotation
of the white dwarf with the orbital period.
In systems with weaker magnetic fields, disks can
form but are truncated by the magnetic field in
its inner part.  This condition is usually met in
intermediate polars (IPs) in which the white dwarf 
rotates asynchronously with the orbital period
[for a review of IPs, see e.g. \citet{pat94ipreview}].

   Several DNe have been confirmed to be IPs; especially
notable are GK Per \citep{wat85gkperspin}\footnote{
   Although GK Per is usually considered as a classical nova,
   it also shows dwarf nova-type outbursts
   (e.g. \cite{bia86gkper}).
}
and DO Dra (\cite{pat92dodra}; \cite{pat93dodraXray}).\footnote{
   Also referred to as YY Dra.
}  They are both
objects above the period gap and are not expected to
develop tidal instability.  In recent years,
several DNe below the period gap have been identified or 
proposed to be IPs, including CC Scl, the subject of this paper.
The inner disk is supposed to be truncated in such systems,
and it may affect the global dynamics of the disk in outburst.
Such a system is expected to provide us insight into
the effect of magnetism in development of outbursts
and superoutbursts, and may eventually help us better
understanding the mechanism of outbursts and superhumps.

\section{CC Sculptoris}

   CC Scl was discovered as a ROSAT source (RX J2315.5$-$3049)
and was optically identified as a dwarf nova, although
outbursts had not been detected \citep{sch00RASSID}.
R. Stubbings visually detected two outbursts in 2000
(vsnet-outburst 245, 810).  During the second outburst,
\citet{ish01j2315} detected likely superhumps with a period
of 0.078~d and amplitudes of $\sim$0.3 mag.
The orbital period, however, was reported to be much
shorter (0.058~d) according to Augusteijn et al.
(2000, vsnet-campaign 544).  \citet{ish01j2315} interpreted
that this discrepancy may be understood if the object
is an IP.  The shorter period was later
confirmed to be the orbital period (\cite{che01ECCV};
\cite{tap04CTCV}).

   During the 2011 superoutburst, \citet{Pdot4} detected
superhumps with a mean period of 0.0600~d.  \citet{Pdot4},
using the least absolute shrinkage and selection operator (Lasso)
method (\cite{lasso}; \cite{kat12perlasso}), demonstrated
that the irregular profiles seen in the superhumps
in this systems are caused by the superposition of
superhumps and orbital modulations.
\citet{wou12ccscl}, partly using the same data as in
\citet{Pdot4}, identified a superhump period
of 1.443~hr (0.0601~d) and also showed that the object
is an IP with a spin period of 0.00450801(6)~d
(389.49~s).

   There was an outburst in 2012 August, which turned out
to be a normal outburst (vsnet-alert 14880, 14892).\footnote{
   VSNET archive can be accessed at (for example, alert messages)
   $<$http://ooruri.kusastro.kyoto-u.ac.jp/pipermail/vsnet-alert/$>$.
}
There was also a normal outburst in 2013 January
(vsnet-alert 15307).

\section{Observations and Analysis}\label{sec:obs}

   The 2014 superoutburst was detected by P. Starr
on July 2 (cvnet-outburst 6019).  The initial peak was actually
a precursor outburst and the main superoutburst followed
five days later (vsnet-alert 17483).  Time-series observations
during this outburst started relatively late and were rather sparse 
compared to the 2011 observation.
Although superhumps were detected,
the object started fading rapidly within three days
(vsnet-alert 17491).  We only observed the later part of
the superoutburst.

   The summary of observations, with mean magnitudes, are listed in
table \ref{tab:log}.  The observer's codes are P. Starr (SPE,
50cm telescope, Warrumbungle Observatory),
F.-J. Hambsch (HMB, 40cm telescope in San Pedro de Atacama, Chile) 
and A. Oksanen (OAR, Harlingten Observatory 
50cm Planewave telescope in San Pedro 
de Atacama, Chile).

\begin{table*}
\caption{Log of observations.}\label{tab:log}
\begin{center}
\begin{tabular}{ccccccc}
\hline
Start\commenta & End\commenta & Mean Mag. & Error & $N$\commentb & Observer & Filter\commentc \\
\hline
56841.2786 & 56841.2833 & 15.089 & 0.028 & 5 & SPE & CV \\
56842.2787 & 56842.2834 & 15.929 & 0.022 & 5 & SPE & CV \\
56843.2777 & 56843.2823 & 16.281 & 0.039 & 5 & SPE & CV \\
56846.2485 & 56846.2658 & 13.604 & 0.037 & 8 & SPE & CV \\
56847.1636 & 56847.1778 & 13.512 & 0.023 & 10 & SPE & CV \\
56848.3128 & 56848.3212 & 13.886 & 0.017 & 6 & SPE & CV \\
56849.1717 & 56849.2915 & 14.135 & 0.009 & 195 & SPE & CV \\
56850.0837 & 56850.1978 & 14.129 & 0.010 & 174 & SPE & V \\
56851.7989 & 56851.9274 & 14.454 & 0.032 & 39 & HMB & CV \\
56852.1178 & 56852.2399 & 15.229 & 0.019 & 66 & SPE & V \\
56852.7910 & 56852.9166 & 16.477 & 0.007 & 296 & OAR & CV \\
56852.7959 & 56852.9276 & 16.349 & 0.027 & 39 & HMB & CV \\
56853.6983 & 56853.9167 & 16.415 & 0.007 & 286 & OAR & CV \\
56853.7932 & 56853.9282 & 16.439 & 0.023 & 43 & HMB & CV \\
56854.7914 & 56854.9279 & 16.526 & 0.018 & 52 & HMB & CV \\
56854.8147 & 56854.9167 & 16.729 & 0.009 & 131 & OAR & CV \\
56855.7348 & 56855.9164 & 16.573 & 0.011 & 236 & OAR & CV \\
56855.7885 & 56855.9254 & 16.641 & 0.018 & 52 & HMB & CV \\
56856.7999 & 56856.9169 & 16.671 & 0.011 & 151 & OAR & CV \\
56856.8046 & 56856.9263 & 16.675 & 0.024 & 39 & HMB & CV \\
56857.7322 & 56857.9164 & 16.631 & 0.007 & 240 & OAR & CV \\
56857.7851 & 56857.9256 & 16.621 & 0.017 & 57 & HMB & CV \\
56858.7823 & 56858.9328 & 16.703 & 0.016 & 61 & HMB & CV \\
56859.2866 & 56859.2955 & 16.749 & 0.063 & 10 & SPE & V \\
56859.7795 & 56859.9320 & 16.607 & 0.021 & 62 & HMB & CV \\
56860.2678 & 56860.2766 & 16.584 & 0.049 & 10 & SPE & V \\
56860.7767 & 56860.9318 & 16.647 & 0.018 & 63 & HMB & CV \\
56861.2658 & 56861.2668 & 16.557 & 0.035 & 2 & SPE & CV \\
56861.7739 & 56861.8067 & 16.745 & 0.048 & 10 & HMB & CV \\
56862.7711 & 56862.9317 & 16.748 & 0.017 & 65 & HMB & CV \\
56863.7685 & 56863.9310 & 16.483 & 0.024 & 66 & HMB & CV \\
56864.7657 & 56864.9315 & 16.874 & 0.014 & 56 & HMB & CV \\
\hline
  \multicolumn{7}{l}{\commenta BJD$-$2400000.} \\
  \multicolumn{7}{l}{\commentb Number of observations.} \\
  \multicolumn{7}{l}{\commentc CV indicates unfiltered observations.} \\
\end{tabular}
\end{center}
\end{table*}

\addtocounter{table}{-1}
\begin{table*}
\caption{Log of observations (continued).}
\begin{center}
\begin{tabular}{ccccccc}
\hline
Start\commenta & End\commenta & Mean Mag. & Error & $N$\commentb & Observer & Filter\commentc \\
\hline
56865.0475 & 56865.0485 & 16.603 & 0.157 & 2 & SPE & CV \\
56865.7629 & 56865.9315 & 16.932 & 0.015 & 57 & HMB & CV \\
56866.0201 & 56866.0211 & 16.897 & 0.052 & 2 & SPE & CV \\
56866.7601 & 56866.9296 & 16.887 & 0.018 & 58 & HMB & CV \\
56867.0184 & 56867.0194 & 16.911 & 0.014 & 2 & SPE & CV \\
56867.7575 & 56867.9291 & 16.769 & 0.018 & 58 & HMB & CV \\
56868.1102 & 56868.1111 & 16.483 & 0.150 & 2 & SPE & CV \\
56868.7545 & 56868.9282 & 16.881 & 0.013 & 60 & HMB & CV \\
56869.1737 & 56869.1747 & 16.699 & 0.120 & 2 & SPE & CV \\
56869.7518 & 56869.9291 & 16.923 & 0.017 & 60 & HMB & CV \\
56869.8215 & 56869.9163 & 16.897 & 0.011 & 126 & OAR & CV \\
56870.0318 & 56870.0327 & 16.834 & 0.063 & 2 & SPE & CV \\
56870.7490 & 56870.9291 & 16.912 & 0.017 & 61 & HMB & CV \\
56871.1968 & 56871.1978 & 16.995 & 0.051 & 2 & SPE & CV \\
56871.7462 & 56871.9293 & 16.923 & 0.017 & 62 & HMB & CV \\
56871.7568 & 56871.9168 & 16.903 & 0.007 & 208 & OAR & CV \\
56872.0608 & 56872.0618 & 16.598 & 0.039 & 2 & SPE & CV \\
56872.7434 & 56872.9258 & 16.981 & 0.015 & 62 & HMB & CV \\
56873.0342 & 56873.0352 & 16.754 & 0.129 & 2 & SPE & CV \\
56873.7406 & 56873.9282 & 16.924 & 0.015 & 64 & HMB & CV \\
56874.0589 & 56874.0589 & 16.288 & -- & 1 & SPE & CV \\
56874.7386 & 56874.9258 & 16.976 & 0.017 & 63 & HMB & CV \\
56875.0108 & 56875.0118 & 16.664 & 0.048 & 2 & SPE & CV \\
56875.7358 & 56875.9263 & 16.995 & 0.015 & 64 & HMB & CV \\
56875.9991 & 56876.0001 & 16.700 & 0.042 & 2 & SPE & CV \\
56876.7330 & 56876.9288 & 16.939 & 0.016 & 66 & HMB & CV \\
56877.0326 & 56877.0336 & 17.038 & 0.233 & 2 & SPE & CV \\
56878.0904 & 56878.0914 & 17.101 & 0.201 & 2 & SPE & CV \\
56879.1822 & 56879.1832 & 16.848 & 0.042 & 2 & SPE & CV \\
56879.7247 & 56879.9118 & 16.991 & 0.024 & 60 & HMB & CV \\
56880.7218 & 56880.9111 & 17.071 & 0.034 & 64 & HMB & CV \\
56881.0641 & 56881.0651 & 16.940 & 0.046 & 2 & SPE & CV \\
56882.9707 & 56882.9717 & 17.049 & 0.045 & 2 & SPE & CV \\
\hline
  \multicolumn{7}{l}{\commenta BJD$-$2400000.} \\
  \multicolumn{7}{l}{\commentb Number of observations.} \\
  \multicolumn{7}{l}{\commentc CV indicates unfiltered observations.} \\
\end{tabular}
\end{center}
\end{table*}

   In period analysis, we used
the 2011 observation \citep{Pdot4} and the 2012 and 2013
observations from the public data in the AAVSO database\footnote{
   $<$http://www.aavso.org/data-download$>$.
} in addition to the 2014 observations.
We also used the Catalina Real-time Transient Survey 
(CRTS; \cite{CRTS})\footnote{
   $<$http://nesssi.cacr.caltech.edu/catalina/$>$.
} for determining the orbital period.

   We adjusted the zero-points between observers and all
observations were converted to Barycentric Julian Days (BJD).
In making period analysis or obtaining phase-averaged
light curves, we removed the long-term trends by using
locally-weighted polynomial regression 
(LOWESS: \cite{LOWESS}).

   We used phase dispersion minimization (PDM; \cite{PDM})
for period analysis and 1$\sigma$ errors for the PDM analysis
was estimated by the methods of \citet{fer89error} and \citet{Pdot2}.
We analyzed 100 samples which randomly contain 50\% of
observations, and performed PDM analysis for these samples.
The bootstrap result is shown as a form of 90\% confidence intervals
in the resultant $\theta$ statistics.

\section{Results}\label{sec:result}

\subsection{Overall Light Curve of Outburst}\label{sec:overall}

\begin{figure*}
  \begin{center}
    \FigureFile(130mm,100mm){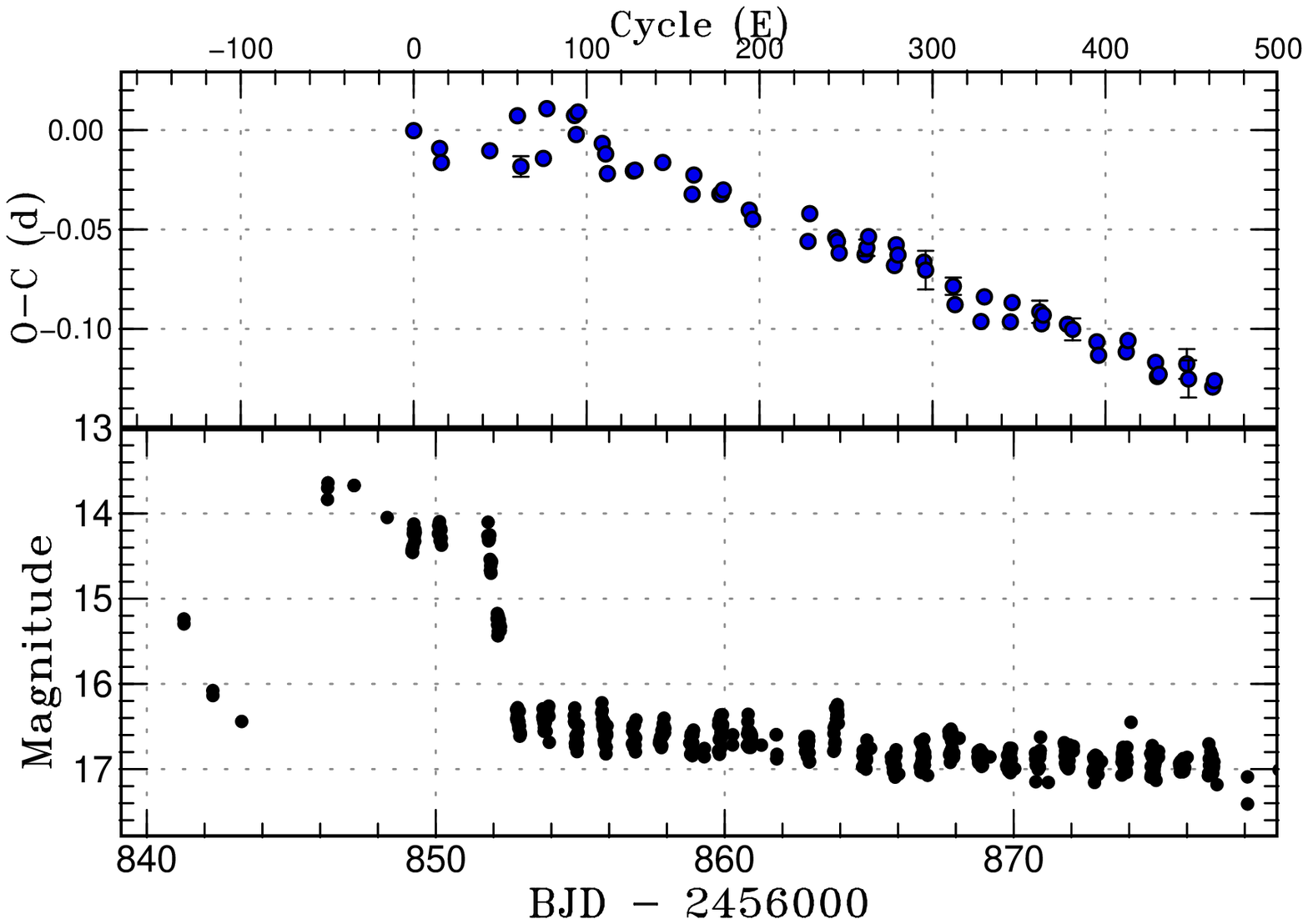}
  \end{center}
  \caption{$O-C$ diagram of superhumps in CC Scl (2014).
     (Upper): $O-C$ diagram.  A period of 0.05986~d
     was used to draw this figure.  A longer superhump period
     (stage B) was observed during the superoutburst.
     After fading from the superoutburst plateau, there was
     a transition to a shorter constant period (stage C superhumps).
     There was no phase jump between them.
     (Lower): Light curve.  The observations were binned to 0.01~d.
     A precursor outburst was clearly detected.}
  \label{fig:ccsclhumpall}
\end{figure*}

   The overall light curve (lower panel of figure \ref{fig:ccsclhumpall})
clearly indicates the presence of a precursor outburst, followed
by temporary fading for at least two days.
Although the start of the main superoutburst was not covered by
observations, its duration was less than 9~d.
This duration is shorter than typical ones (10--14~d)
in SU UMa-type dwarf novae with short orbital periods
(e.g. \cite{nog98swuma}; \cite{bab00v1028cyg}).
There was no post-superoutburst rebrightening.

   The mean fading rate from the precursor outburst
for the first two days was 0.84(3) mag d$^{-1}$, which is
somewhat slower than fading rates of normal outbursts in
SU UMa-type dwarf novae.
The mean fading rate of the plateau stage of
the superoutburst was 0.11(1) mag d$^{-1}$, which is
also typical for an SU UMa-type dwarf nova
other than candidate period bouncers \citep{Pdot5}.
The mean fading rate during the rapid fading
from the plateau phase was 1.69(2) mag d$^{-1}$.
Slow fading at a rate of 0.021(1) mag d$^{-1}$
continued for more than 20~d after the rapid fading.
This feature is frequently seen in SU UMa-type dwarf novae
with infrequent outbursts or in WZ Sge-type dwarf novae.

\subsection{Superhumps during Superoutburst and Early Post-Superoutburst}
\label{sec:superhump}

\begin{figure}
  \begin{center}
    \FigureFile(88mm,110mm){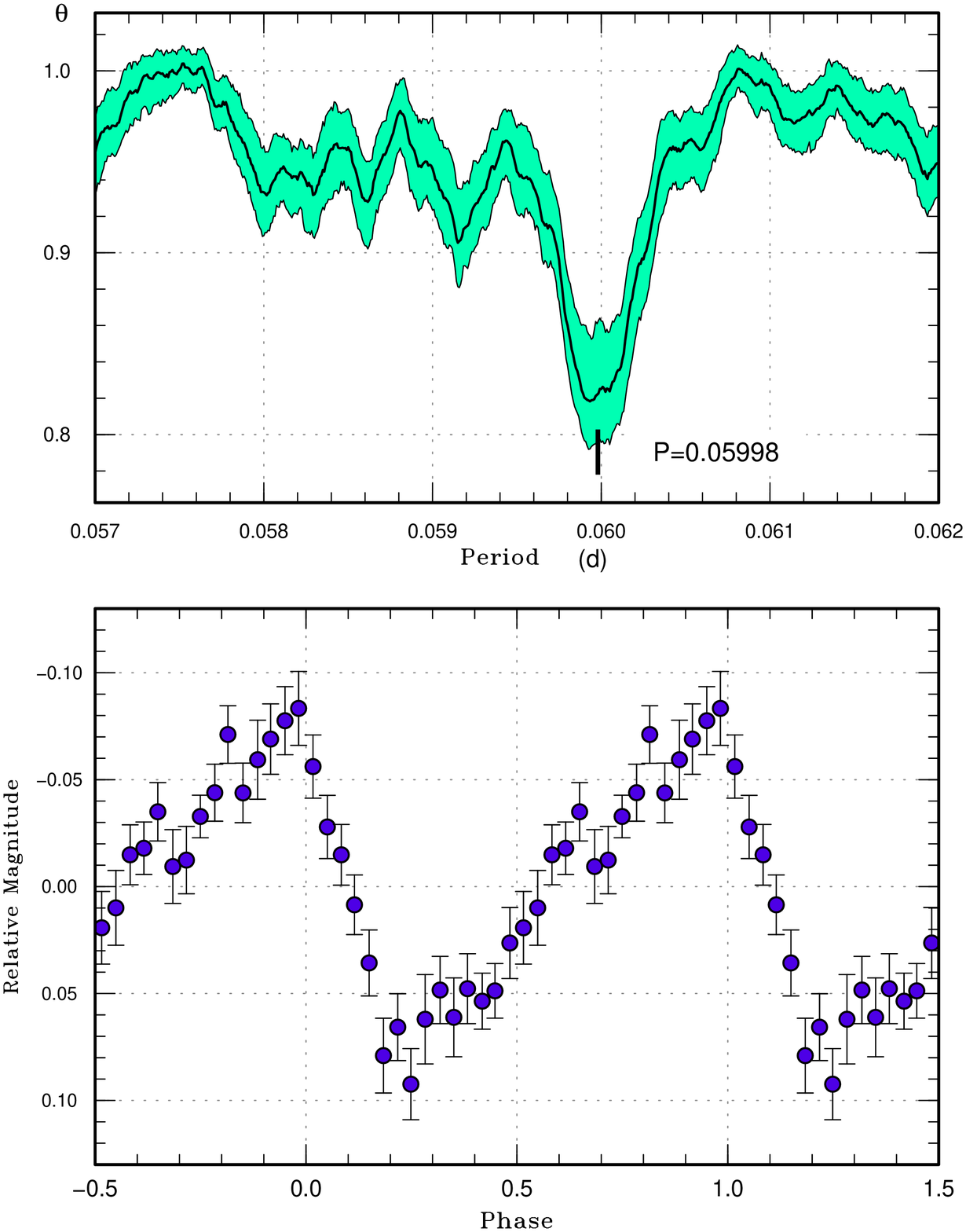}
  \end{center}
  \caption{Superhumps in CC Scl during superoutburst and three nights
     just after the superoutburst (2014).
     (Upper): PDM analysis.
     The 90\% confidence intervals by the bootstraping method
     (see text for the details) is shown by two curves above
     and below the central curve.
     (Lower): Phase-averaged profile by the superhump period
     of 0.05998~d.}
  \label{fig:ccsclshpdm1}
\end{figure}

   The times of superhump maxima were determined by the template
fitting method as described in \citet{Pdot}.
The results during the superoutburst are listed in table \ref{tab:ccscloc2014}.
Due to the limited coverage during the superoutburst,
the statistics were rather poor.  As shown later
(subsection \ref{sec:superhumpstage}), this superhump period
persisted up to three days after the rapid fading
from the superoutburst plateau.  Figure \ref{fig:ccsclshpdm1}
shows the PDM analysis and phase-averaged profile during
the superoutburst and the three nights just after
the superoutburst.  The best period by the PDM method
was 0.05998(2)~d.

\begin{table}
\caption{Superhump maxima of CC Scl (2014)}\label{tab:ccscloc2014}
\begin{center}
\begin{tabular}{rp{50pt}p{30pt}r@{.}lcr}
\hline
$E$ & max\commenta & error & \multicolumn{2}{c}{$O-C$\commentb} & phase\commentc & $N$\commentd \\
\hline
0 & 56849.2357 & 0.0014 & 0&0076 & 0.38 & 81 \\
15 & 56850.1246 & 0.0015 & $-$0&0014 & 0.56 & 79 \\
16 & 56850.1774 & 0.0011 & $-$0&0085 & 0.46 & 64 \\
44 & 56851.8594 & 0.0021 & $-$0&0025 & 0.18 & 15 \\
60 & 56852.8348 & 0.0011 & 0&0151 & 0.83 & 126 \\
62 & 56852.9290 & 0.0052 & $-$0&0104 & 0.44 & 51 \\
\hline
  \multicolumn{7}{l}{\commenta BJD$-$2400000.} \\
  \multicolumn{7}{l}{\commentb Against max $= 2456849.2281 + 0.059859 E$.} \\
  \multicolumn{7}{l}{\commentc Orbital phase.} \\
  \multicolumn{7}{l}{\commentd Number of points used to determine the maximum.} \\
\end{tabular}
\end{center}
\end{table}

\subsection{Post-Superoutburst Superhumps}\label{sec:postsuper}

\begin{figure}
  \begin{center}
    \FigureFile(88mm,110mm){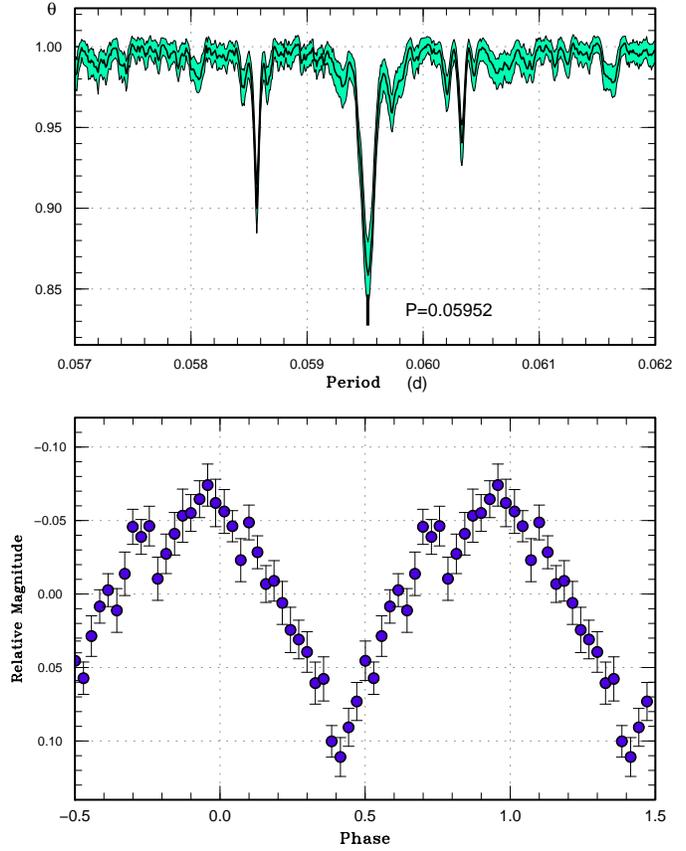}
  \end{center}
  \caption{Superhumps in CC Scl in the postsuperoutburst phase (2014).
     (Upper): PDM analysis.  The sharp signal at
     0.05857~d is the orbital period.
     The 90\% confidence intervals by the bootstraping method
     (see text for the details) is shown by two curves above
     and below the central curve.
     (Lower): Phase-averaged profile by the superhump period
     of 0.059523~d.}
  \label{fig:ccsclpostshpdm}
\end{figure}

   During the post-superoutburst stage except the initial
three nights, a PDM analysis yielded a stable period of 0.059523(6)~d
(figure \ref{fig:ccsclpostshpdm}).
The period is considerably shorter than the period
of superhumps during the superoutburst.
The times of post-superoutburst maxima (including the initial
three nights) are listed in table
\ref{tab:ccscloc2014b}.

\begin{table}
\caption{Superhump maxima of CC Scl (2014) (post-superoutburst)}\label{tab:ccscloc2014b}
\begin{center}
\begin{tabular}{rp{50pt}p{30pt}r@{.}lcr}
\hline
$E$ & max\commenta & error & \multicolumn{2}{c}{$O-C$\commentb} & phase\commentc & $N$\commentd \\
\hline
0 & 56853.7113 & 0.0033 & $-$0&0174 & 0.80 & 45 \\
2 & 56853.8560 & 0.0022 & 0&0084 & 0.27 & 75 \\
18 & 56854.8103 & 0.0009 & 0&0104 & 0.56 & 38 \\
19 & 56854.8606 & 0.0024 & 0&0012 & 0.42 & 75 \\
20 & 56854.9318 & 0.0021 & 0&0128 & 0.64 & 36 \\
34 & 56855.7540 & 0.0009 & 0&0019 & 0.68 & 51 \\
36 & 56855.8685 & 0.0009 & $-$0&0027 & 0.63 & 78 \\
37 & 56855.9184 & 0.0019 & $-$0&0123 & 0.48 & 36 \\
52 & 56856.8176 & 0.0008 & $-$0&0058 & 0.84 & 54 \\
53 & 56856.8779 & 0.0007 & $-$0&0050 & 0.87 & 79 \\
69 & 56857.8395 & 0.0010 & 0&0043 & 0.29 & 82 \\
86 & 56858.8412 & 0.0013 & $-$0&0058 & 0.39 & 22 \\
87 & 56858.9107 & 0.0024 & 0&0043 & 0.58 & 20 \\
102 & 56859.7990 & 0.0012 & $-$0&0002 & 0.74 & 13 \\
103 & 56859.8589 & 0.0015 & 0&0002 & 0.76 & 20 \\
104 & 56859.9208 & 0.0015 & 0&0026 & 0.82 & 17 \\
119 & 56860.8086 & 0.0018 & $-$0&0024 & 0.98 & 17 \\
121 & 56860.9236 & 0.0032 & $-$0&0063 & 0.95 & 16 \\
153 & 56862.8281 & 0.0018 & $-$0&0064 & 0.46 & 21 \\
154 & 56862.9019 & 0.0023 & 0&0079 & 0.72 & 20 \\
169 & 56863.7877 & 0.0016 & 0&0010 & 0.85 & 12 \\
170 & 56863.8457 & 0.0014 & $-$0&0006 & 0.84 & 21 \\
171 & 56863.8997 & 0.0022 & $-$0&0060 & 0.76 & 20 \\
186 & 56864.7968 & 0.0031 & $-$0&0017 & 0.08 & 14 \\
187 & 56864.8601 & 0.0041 & 0&0021 & 0.16 & 16 \\
188 & 56864.9256 & 0.0025 & 0&0081 & 0.28 & 12 \\
203 & 56865.8090 & 0.0022 & $-$0&0013 & 0.36 & 17 \\
204 & 56865.8792 & 0.0017 & 0&0095 & 0.56 & 16 \\
205 & 56865.9340 & 0.0031 & 0&0047 & 0.49 & 9 \\
220 & 56866.8283 & 0.0018 & 0&0062 & 0.76 & 17 \\
221 & 56866.8841 & 0.0097 & 0&0026 & 0.72 & 17 \\
237 & 56867.8338 & 0.0043 & 0&0000 & 0.93 & 16 \\
\hline
  \multicolumn{7}{l}{\commenta BJD$-$2400000.} \\
  \multicolumn{6}{l}{\commentb Against max $= 2456853.7286 + 0.059515 E$.} \\
  \multicolumn{7}{l}{\commentc Orbital phase.} \\
  \multicolumn{7}{l}{\commentd Number of points used to determine the maximum.} \\
\end{tabular}
\end{center}
\end{table}

\addtocounter{table}{-1}
\begin{table}
\caption{Superhump maxima of CC Scl (2014) (post-superoutburst, continued)}
\begin{center}
\begin{tabular}{rp{50pt}p{30pt}r@{.}lcr}
\hline
$E$ & max\commenta & error & \multicolumn{2}{c}{$O-C$\commentb} & phase\commentc & $N$\commentd \\
\hline
238 & 56867.8844 & 0.0037 & $-$0&0089 & 0.80 & 16 \\
253 & 56868.7738 & 0.0014 & $-$0&0123 & 0.98 & 11 \\
255 & 56868.9059 & 0.0027 & 0&0008 & 0.24 & 16 \\
270 & 56869.7912 & 0.0036 & $-$0&0066 & 0.35 & 16 \\
271 & 56869.8608 & 0.0015 & 0&0035 & 0.54 & 78 \\
287 & 56870.8139 & 0.0056 & 0&0044 & 0.82 & 17 \\
288 & 56870.8678 & 0.0023 & $-$0&0013 & 0.73 & 16 \\
289 & 56870.9320 & 0.0037 & 0&0034 & 0.83 & 9 \\
303 & 56871.7654 & 0.0012 & 0&0036 & 0.06 & 47 \\
306 & 56871.9424 & 0.0055 & 0&0021 & 0.08 & 12 \\
320 & 56872.7742 & 0.0016 & 0&0006 & 0.29 & 14 \\
321 & 56872.8273 & 0.0019 & $-$0&0058 & 0.19 & 16 \\
337 & 56873.7867 & 0.0029 & 0&0014 & 0.57 & 17 \\
338 & 56873.8523 & 0.0038 & 0&0075 & 0.69 & 16 \\
354 & 56874.7991 & 0.0024 & 0&0020 & 0.86 & 17 \\
355 & 56874.8519 & 0.0024 & $-$0&0047 & 0.76 & 16 \\
356 & 56874.9128 & 0.0029 & $-$0&0034 & 0.80 & 15 \\
372 & 56875.8758 & 0.0075 & 0&0074 & 0.24 & 17 \\
373 & 56875.9281 & 0.0094 & 0&0002 & 0.14 & 10 \\
387 & 56876.7621 & 0.0027 & 0&0010 & 0.38 & 13 \\
388 & 56876.8251 & 0.0022 & 0&0045 & 0.45 & 16 \\
438 & 56879.7918 & 0.0050 & $-$0&0046 & 0.11 & 16 \\
454 & 56880.7411 & 0.0041 & $-$0&0075 & 0.32 & 11 \\
455 & 56880.8066 & 0.0055 & $-$0&0015 & 0.44 & 16 \\
\hline
  \multicolumn{7}{l}{\commenta BJD$-$2400000.} \\
  \multicolumn{6}{l}{\commentb Against max $= 2456853.7286 + 0.059515 E$.} \\
  \multicolumn{7}{l}{\commentc Orbital phase.} \\
  \multicolumn{7}{l}{\commentd Number of points used to determine the maximum.} \\
\end{tabular}
\end{center}
\end{table}

\subsection{Orbital Variations}\label{sec:orbital}

   The reported orbital period of 0.05763~d \citep{wou12ccscl}
could not be detected in the analysis of the post-superoutburst
data (figure \ref{fig:ccsclpostshpdm}).
We should note that if 0.05763~d is the true
orbital period, the fractional superhump excess
$\epsilon \equiv P_{\rm SH}/P_{\rm orb}-1$=4.3\%,
using the values in \citet{wou12ccscl}, is
too large for this orbital period.

\begin{figure}
  \begin{center}
    \FigureFile(88mm,70mm){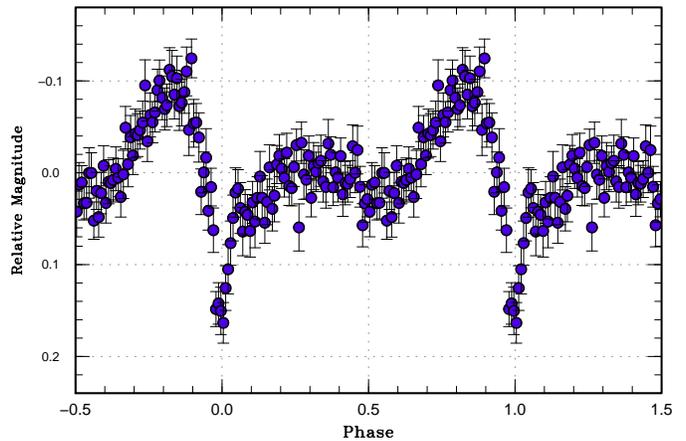}
  \end{center}
  \caption{Mean orbital light curve of CC Scl in quiescence.
      The ephemeris of equation (\ref{equ:ccsclecl}) is used.
      Time-series observations fainter than 16 mag were used.}
  \label{fig:ccsclquiporb}
\end{figure}

\begin{figure}
  \begin{center}
    \FigureFile(88mm,110mm){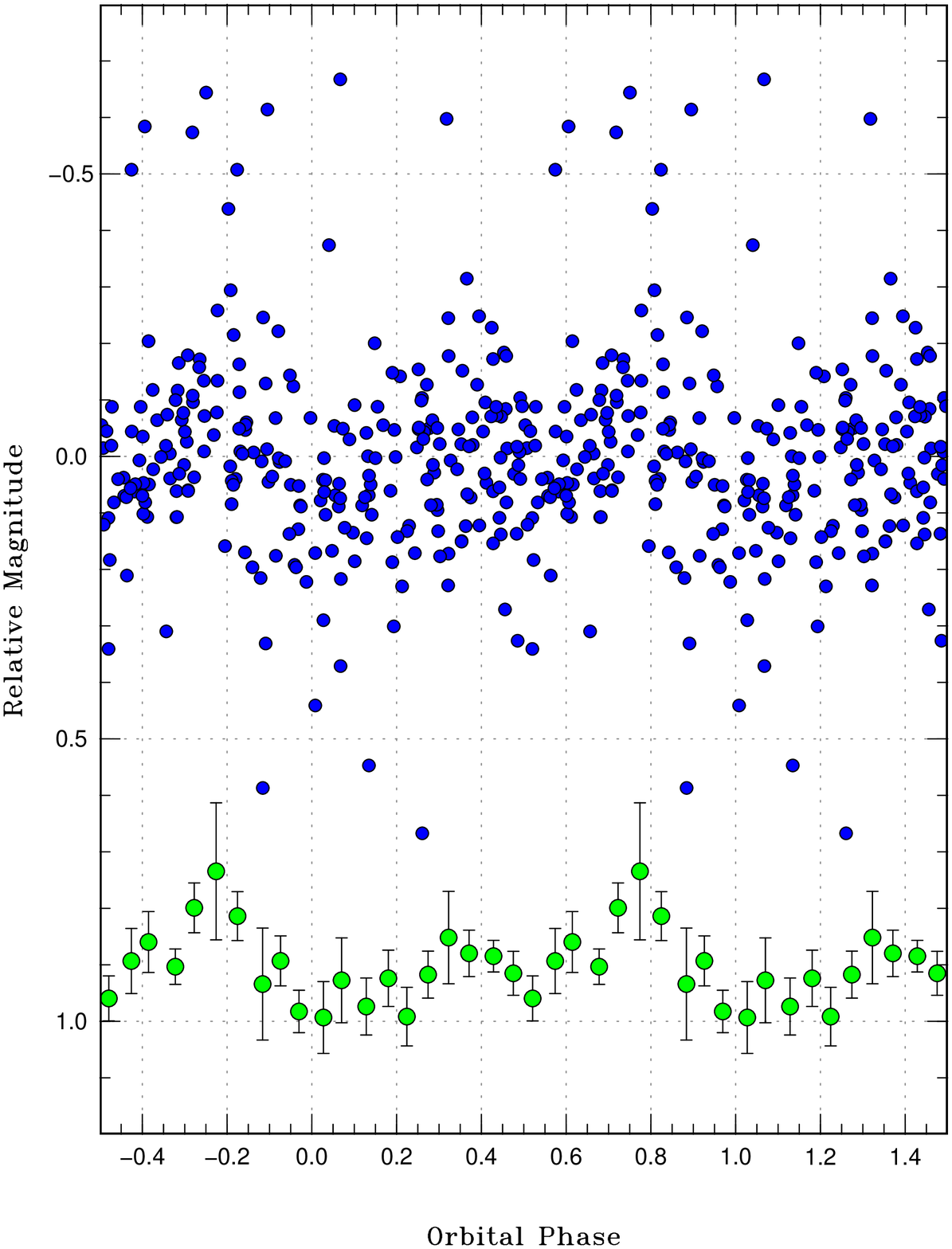}
  \end{center}
  \caption{Orbital light curve of CC Scl from the CRTS data in
      quiescence.  The ephemeris of equation (\ref{equ:ccsclecl})
      is used.  Long-term trends were subtracted.
      Typical errors of the CRTS observations were 0.10--0.18 mag.
      In the lower part of the figure, phase-averaged data to
      20 bins are plotted with larger symbols and 1$\sigma$ errors.}
  \label{fig:ccsclcrts}
\end{figure}

   We alternately propose the orbital period of 0.058566(2)~d 
detected in the PDM analysis of the post-superoutburst observations
(figure \ref{fig:ccsclpostshpdm}).
This value is in good agreement with the one 0.05845~d
by \citet{che01ECCV} and the period [0.0585845(10)~d]
detected during the 2011 observation \citep{Pdot4}.
By adopting this period, the orbital light curve turned
out to show a shallow eclipse with double orbital humps.\footnote{
   A. Oksanen also noticed the presence of eclipse-like fading
   during the 2013 observations in quiescence.
}
Figure \ref{fig:ccsclquiporb} represents quiescent time-series
observations for 2011--2014 when the magnitude was fainter
than 16 (these observations include post-superoutburst
observations).
It is worth noting that
\citet{che01ECCV} correctly referred to a dip observed
in the light curve as a possible eclipse.
The ephemeris of eclipses using all the available data
(2011, 2012, 2014 outburst and post-outburst observations)
and the CRTS data was determined by Markov-chain Monte Carlo (MCMC)
method, which was introduced in \citet{Pdot4}, as follows:
\begin{equation}
{\rm Min(BJD)} = 2456668.00638(9) + 0.058567233(8) E .
\label{equ:ccsclecl}
\end{equation}
Since the times of individual eclipses are difficult to
determine, we instead give the mean epoch [BJD 2456863.5624(1)]
of the eclipse observations after the 2014 superoutburst.
The phase plot of the CRTS observations is also shown
in figure \ref{fig:ccsclcrts}.  Although eclipses
are not very clear in the CRTS data, orbital modulations
having a hump maximum around phase 0.8 were recorded
and the overall appearance appears to be consistent with
the 2014 post-superoutburst observations in quiescence.

   The quiescent orbital profile resembles those of
low mass-transfer rate objects such as WZ Sge-type dwarf novae
[see e.g. WZ Sge and AL Com \citep{pat96alcom}, V455 And
(\cite{ara05v455and}; \cite{Pdot}), V386 Ser \citep{muk10v386ser},
EZ Lyn (\cite{kat09j0804}; \cite{zha13ezlyn}), BW Scl
(\cite{aug97bwscl}; \cite{Pdot4})], although the double-wave
orbital humps are less clear in CC Scl.
A classical interpretation assuming a semi-transparent accretion
disk allowing the light from the hot spot to escape in two directions
\citep{ski00wzsge} would be a viable interpretation.

\begin{figure}
  \begin{center}
    \FigureFile(88mm,70mm){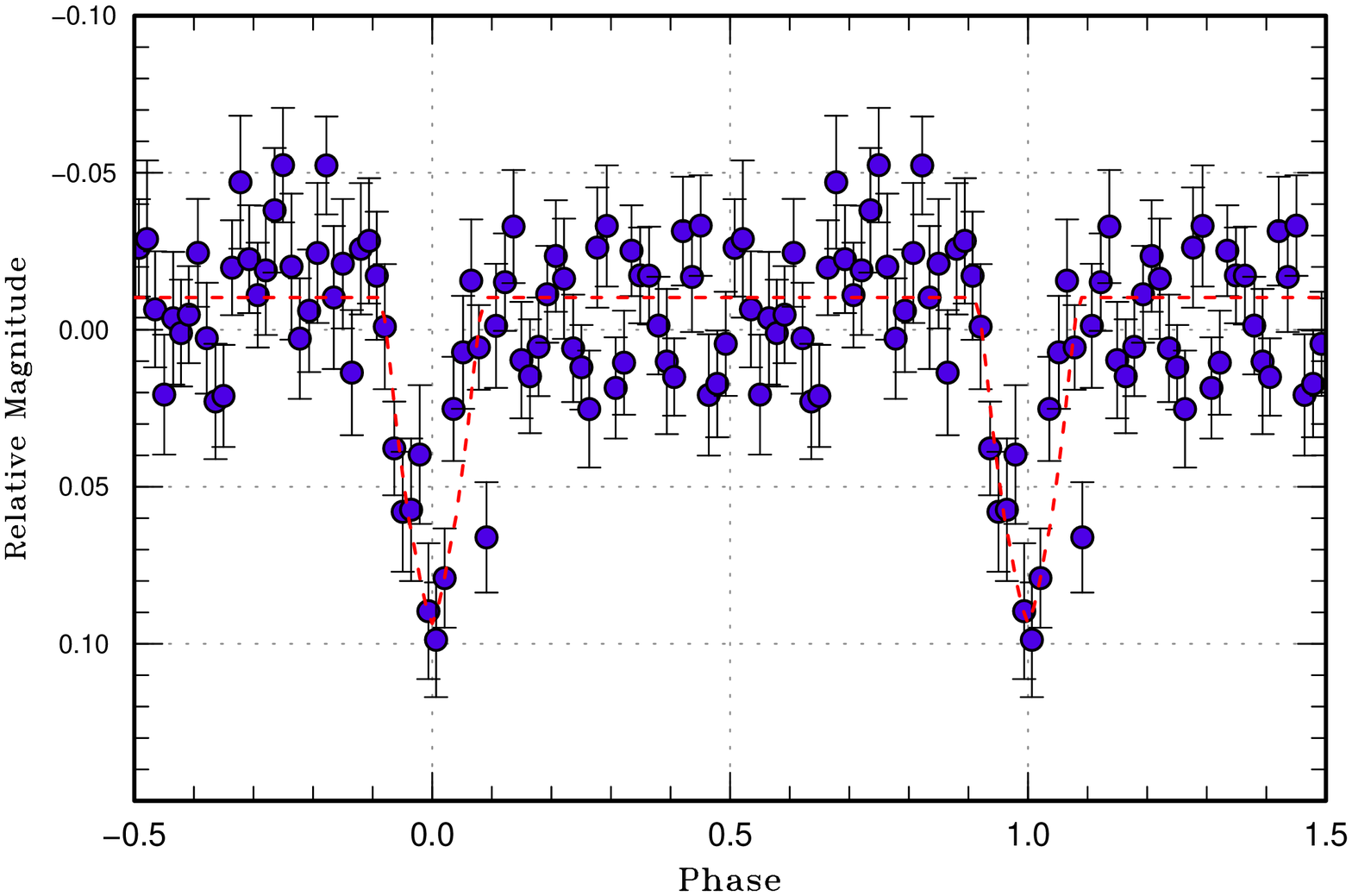}
  \end{center}
  \caption{Mean orbital light curve of CC Scl in outburst.
      The ephemeris of equation (\ref{equ:ccsclecl}) is used.
      Time-series observations during the 2011, 2012 and 2014
      outbursts were used.  The dashed line represents a model
      with $q$=0.072, $i$=\timeform{D80.6} and disk radius
      of 0.41$A$ (see subsection \ref{sec:parameters}).}
  \label{fig:ccscloutporb}
\end{figure}

    Although eclipses became less apparent in outburst,
they continued to be present (figure \ref{fig:ccscloutporb}).
Orbital humps almost disappeared in outburst.

\section{Discussion}\label{sec:discussion}

\subsection{Identification of Superhump Stages}\label{sec:superhumpstage}

   By using the new orbital period, the fractional superhumps
excesses are found to be within a reasonable region:
2.6\% for the 2011 data, 2.4\% for the 2014 data
during superoutburst and 1.6\% for the post-superoutburst
data in 2014, respectively.

   In most SU UMa-type dwarf novae, shortening of
the superhump period is not usually observed following
the rapid fading from the superoutburst, and the period
after the fading is usually the same as that of stage C
superhumps, which are late-stage superhumps with
almost constant periods [for the definition of
superhump stages, see \citet{Pdot}].
Such shortening of the superhump period immediately 
following the rapid decline apparently is unique to CC Scl.

   The difference of fractional superhump excesses
during the superoutburst and post-superoutburst is
0.8--1.0\%, which is similar to what is usually
observed between stage B and C superhumps
\citep{Pdot}.  We therefore identify the superhumps
during the post-superoutburst phase to be stage C
superhumps and those during the superoutburst
to be stage B superhumps, respectively.
There was no phase jump between these stages
(see upper panel of figure \ref{fig:ccsclhumpall}).
There was no evidence of ``traditional'' late superhumps,
in which an $\sim$0.5 phase jump is observed
(e.g. \cite{vog83lateSH}).  This finding strengthens our
interpretation of the period change as stage B--C
transition [see \citet{Pdot} for the lack of a phase jump
between stages B and C].

   Since the duration of the 2014 superoutburst was relatively
short (less than 9~d excluding the precursor part;
the duration of the 2011 superoutburst was less constrained but
was shorter than 11~d),
it may be possible that the superoutburst ended
earlier than in other SU UMa-type dwarf novae
and the stage B--C transition was consequently recorded 
in the later phase of the outburst than in other
SU UMa-type dwarf novae.  Such early termination of
the outburst can be reasonably explained assuming
that the inner part of the disk is drained by
the magnetic field of the white dwarf \citep{wou12ccscl}.
Since stage B--C transition was not apparently affected
by this effect, we can suggest that stage C superhumps
originate from the outer part of the accretion disk,
rather than the inner part.  Further observations
of superhumps in such systems may shed light on
the origin of still unresolved stage C superhumps.

\subsection{Orbital Parameters}\label{sec:parameters}

   Since we did not observe stage A superhumps,
we could not directly apply the modern method
of estimating the mass ratio ($q$) from the fractional
superhump excess \citep{kat13qfromstageA}.
We can, however, constrain $q$ using the post-superoutburst
superhumps.  This method was introduced in
\citet{kat13j1222}.  We repeat the essence of the method
for clarity.

   The dynamical precession rate, $\omega_{\rm dyn}$
in the disk can be expressed by (see, \cite{hir90SHexcess}):
\begin{equation}
\label{equ:precession}
\omega_{\rm dyn}/\omega_{\rm orb} = Q(q) R(r),
\end{equation}
where $\omega_{\rm orb}$ and $r$ are the angular orbital frequency
and the dimensionless radius measured in units of the binary 
separation $A$.  The dependence on $q$ and $r$ are
\begin{equation}
\label{equ:qpart}
Q(q) = \frac{1}{2}\frac{q}{\sqrt{1+q}},
\end{equation}
and
\begin{equation}
\label{equ:rpart}
R(r) = \frac{1}{2}\frac{1}{\sqrt{r}} b_{3/2}^{(1)}(r),
\end{equation}
where
$\frac{1}{2}b_{s/2}^{(j)}$ is the Laplace coefficient
\begin{equation}
\label{equ:laplace}
\frac{1}{2}b_{s/2}^{(j)}(r)=\frac{1}{2\pi}\int_0^{2\pi}\frac{\cos(j\phi)d\phi}
{(1+r^2-2r\cos\phi)^{s/2}},
\end{equation}
This $\omega_{\rm dyn}/\omega_{\rm orb}$
is equivalent to the fractional superhump excess (in frequency)
$\epsilon^* \equiv 1-P_{\rm orb}/P_{\rm SH}$
and it is related to the conventional fractional superhump excess 
(in period) $\epsilon$
by a relation $\epsilon^*=\epsilon/(1+\epsilon)$.
This dynamical precession rate is considered to be equal
to the observed $\epsilon^*$ when the pressure effect
can be ignored.  This condition is achieved either if
the superhumps are confined to the region of the 3:1
resonance (stage A superhumps) or the disk is cold such as
in a state of post-superoutburst superhumps
\citep{osa13v344lyrv1504cyg}.

   We can express fractional superhump excesses
(in frequency unit) of post-superoutburst
superhumps as follows:
\begin{equation}
\label{equ:epspost}
\epsilon^*({\rm post}) = Q(q) R(r_{\rm post}),
\end{equation}
where
$\epsilon^*({\rm post})$ and $r_{\rm post}$ are the fractional
superhump excess and disk radius immediately after
the outburst, respectively.

   In various SU UMa-type objects other than WZ Sge-type
dwarf novae with multiple rebrightenings, the value
of $r_{\rm post}$ has been experimentally known to be
in a narrow region 0.37--0.38$A$, where $A$ is
the binary separation \citep{kat13qfromstageA}.
By assuming this $r_{\rm post}$
in CC Scl, we can estimate $q$=0.072(3) (the error corresponds
to the error of $\epsilon^*$).
This value is within a range of $0.06 < q < 0.09$
in \citet{che01ECCV}, who assumed the mass-radius relation
for a normal lower main-sequence secondary.

   By assuming $q$=0.072, we can constrain the binary
inclination by modeling the eclipse profile in outburst.
As in the section of MASTER OT J005740.99$+$443101.5 in
\citet{Pdot6}, we modeled the eclipse light curve.
We assumed flat and axisymmetric geometry and
a standard disk having a surface luminosity with 
a radial dependence $\propto r^{-3/4}$
(i.e. assuming that we observed the Rayleigh-Jeans
tail of the emission from the hot disk).
The secondary is assumed to fill the Roche lobe.
Although these assumptions on the disk are rough, they
will not seriously affect the results.  In the present case,
this is because the central part of this disk needs 
to be grazingly eclipsed to reproduce the shallow eclipse
in outburst, and the result
is very insensitive to the condition in the outer part
of the disk (either radius or the existence of disk flaring). 
For an optically thick disk with a broad range of
radius 0.33--0.46$A$, an inclination value of $i$=\timeform{80.6D}
best reproduced the observed eclipse depth of 0.11 mag
(figure \ref{fig:ccscloutporb}, in which the case of 0.41$A$
is shown as an example).
The uncertainty in $i$ was less than \timeform{0.5D}.

   The depth of eclipses is deeper in quiescence.
This is likely caused by the contribution from
the hot spot.

\subsection{Spin Modulations}\label{sec:spin}

\begin{figure}
  \begin{center}
    \FigureFile(88mm,110mm){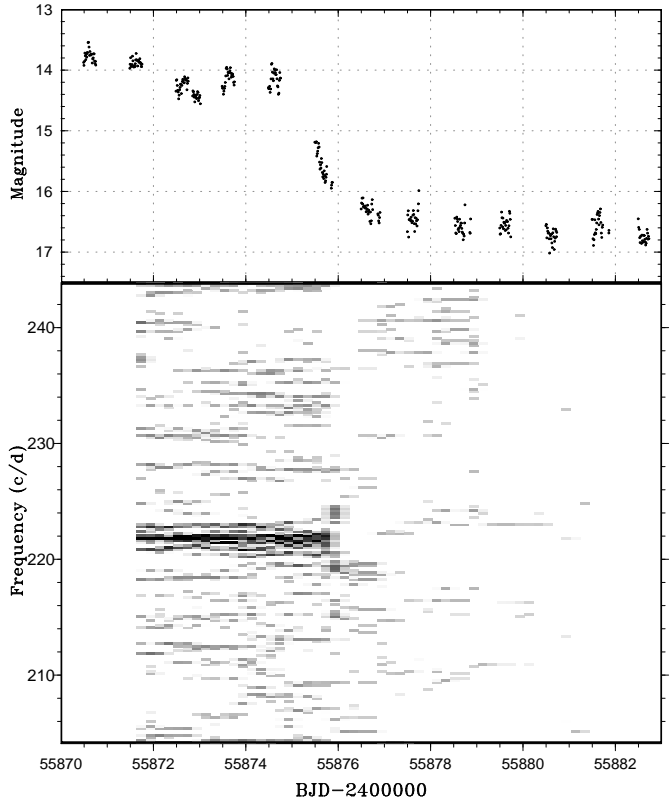}
  \end{center}
  \caption{Lasso 2-D power spectrum analysis of the spin period
     in CC Scl (2011).
     (Upper): Light curve.  The data were binned to 0.01~d.
     (Lower): Result of the lasso analysis ($\log \lambda=-8.5$).
     The frequency 221.8 c/d corresponds to the spin period.
     The spin modulations became strongest around the end of
     the superoutburst, and then decayed quickly.
     The width of the sliding window and the time step used are
     2.5~d and 0.2~d, respectively.
     }
  \label{fig:ccscllasso1}
\end{figure}

\begin{figure}
  \begin{center}
    \FigureFile(88mm,110mm){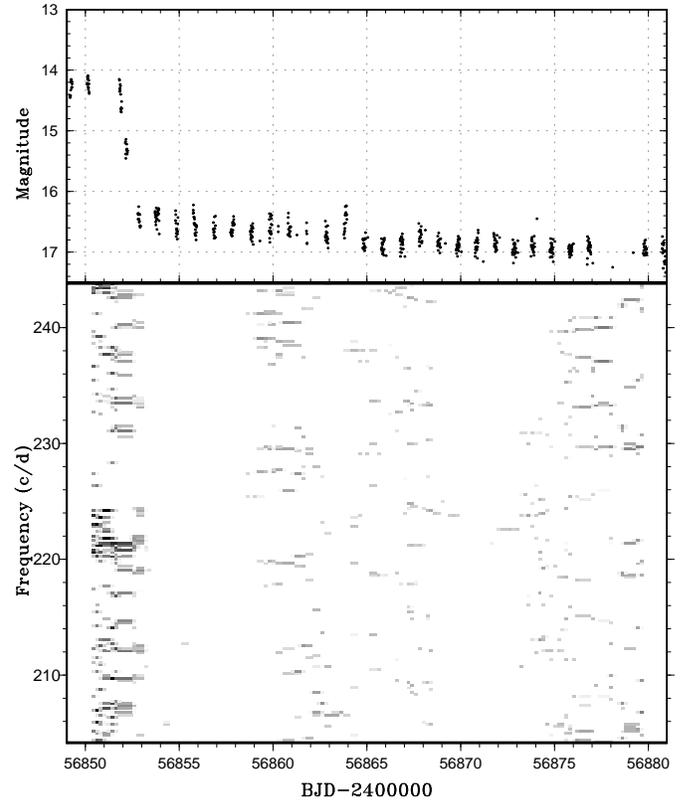}
  \end{center}
  \caption{Lasso 2-D power spectrum analysis of the spin period
     in CC Scl (2014).
     (Upper): Light curve.  The data were binned to 0.01~d.
     (Lower): Result of the lasso analysis ($\log \lambda=-8.5$).
     The frequency 221.8 c/d corresponds to the spin period.
     The spin modulations were detected during the superoutburst
     but were not clearly detected after the ending of
     the superoutburst.  Note that the observation statistics
     was better in 2011.
     The width of the sliding window and the time step used are
     2.5~d and 0.2~d, respectively.
     }
  \label{fig:ccscllasso2}
\end{figure}

   After the detection of the IP spin modulations by
\citet{wou12ccscl}, we re-examined our data in 2011
and examined the present data in 2014.
The spin period could be detected in outburst
observations both in 2011 and 2014.  Using the PDM
method, the 2011 observation
yielded a period of 0.0045076(2)~d (amplitude 0.09 mag)
and the 2014 one yielded 0.0045079(9)~d (amplitude
0.08 mag).  Post-outburst data yielded weaker signals:
0.06 mag in 2011 and 0.04 mag in 2014 in amplitude.
Examples of Lasso 2-D power spectrum analysis 
(cf. \cite{kat13j1924}) are shown in 
figures \ref{fig:ccscllasso1} (the 2011 superoutburst) 
and \ref{fig:ccscllasso2} (the 2014 superoutburst).
Spin modulations in post-superoutburst
stage were not clearly detected in Lasso analysis
since short (2.5~d) windows were used.
Since the system brightness faded by $\sim$3 mag after
the superoutburst, the pulsed flux decreased by a factor
of 15--30 after the outburst.  This phenomenon can be
naturally understood by considering that the pulsed flux reflects
the intensity of the accretion column on the magnetic
pole and that the accretion rate dramatically decreased
after the outburst.  This behavior is consistent with
X-ray observations in \citet{wou12ccscl}.

\subsection{Implication on Disk Instability Model}\label{sec:diskinst}

   Among IPs, V455 And (\cite{ara05v455and};
\cite{sil12v455andGALEX}) has been the only object that
showed a WZ Sge-type superoutburst [although there has been
a claim that WZ Sge (e.g. \cite{war08DNOQPO}) is also
an IP, the situation is less clear].
CC Scl is the first IP that confidently
exhibits a superoutburst of an ordinary SU UMa-type
dwarf nova, rather than an extreme superoutburst of
a WZ Sge-type dwarf nova.
This finding suggests that the basic mechanism causing
ordinary SU UMa-type superoutbursts is not strongly affected
by the magnetism of the white dwarf.
The standard TTI model
\citep{osa89suuma} requires the 3:1 resonance to
trigger a superoutburst.  The radius of the 3:1 resonance
is the outermost achievable
radius in ordinary SU UMa-type dwarf novae.
This radius is larger than any radius in outburst
cycles of normal outbursts, and if the disk is sufficiently
present (or not so strongly truncated) to exhibit normal
outbursts, we can expect that the magnetism will less affect
the disk at the radius of the 3:1 resonance.  We can thus expect
to see superoutbursts if the total angular momentum
accumulates during the cycles of normal outbursts
and the disk radius eventually reaches the 3:1 resonance
on the occasion of an outburst.
This is exactly what is seen in CC Scl, and the observed
behavior is in agreement with the TTI
model considering a partial truncation of the inner disk.

   It is worth noting that a precursor outburst was
also seen in CC Scl.  In the TTI
model, the precursor outburst is the final normal outburst
during which the expansion of the disk brings
its radius to the 3:1 resonance.  Since this phenomenon
takes place also in the outermost part of the disk,
it would be a natural consequence that magnetism does not
strongly affect the appearance of the precursor outburst.

   We should discuss the outburst behavior of other
IPs among dwarf novae below the period gap.
HT Cam is an established IP (\cite{tov98htcam}; \cite{kem02htcam};
\cite{eva05htcam}; \cite{demar05htcam}).
This object shows only brief outbursts with extremely
rapid fading rates (\cite{kem02htcam}; \cite{ish02htcam}).
These outbursts are reported to occur quasi-cyclically,
and can be regarded as a consequence of thermal instability
(\cite{ish02htcam}; \cite{wou12ccscl}).  Despite that
HT Cam should have a mass ratio low enough to develop
the 3:1 resonance, no superoutburst has been recorded. 
As discussed in \citet{wou12ccscl}, the disk in HT Cam may be 
more strongly truncated than in CC Scl, and this may be
responsible for the difference in outburst behavior.
We should continue to see whether HT Cam never shows
a superoutburst.

   FS Aur is an enigmatic object below the period gap.
Although this object has an orbital period of 0.05958096(5)~d
(\cite{tho96Porb}; \cite{neu13fsaur}), only short (normal)
outbursts were observed with short ($\sim$12~d) recurrence times
(\cite{ges89fsaur}; \cite{and91fsaur}).  \citet{neu12fsaur} 
reported that the recurrence time of the normal outbursts
was relatively short and stable (18$\pm$2.5~d). 
VSNET and AAVSO observations since 2010 have basically
confirmed this outburst property.
\citet{neu13fsaur} suggested that this object is an IP.
The light behavior, however, is totally different from
either HT Cam or CC Scl.  FS Aur shows short outbursts
similar to ordinary SU UMa-type normal outbursts,
and does not show extremely rapid fading as seen in
HT Cam [the fading rate during the linear fading part
is reported to be $\sim$0.8 mag d$^{-1}$ \citep{neu12fsaur},
which is an ordinary value for normal outbursts of SU UMa-type
dwarf novae.\footnote{
  Although faster ($\sim$2 mag d$^{-1}$) fading rate was
  reported in the final stage of outbursts \citep{neu12fsaur},
  our examination suggests that
  this feature is not present in many of outbursts recorded
  in the AAVSO database and we do not consider it convincing.
  Since this faster fading rate was recorded only for a brief
  time (less than 1~d), intrinsic erratic variations of the object
  or systematic difference in the zero-point between different
  observers may have affected the conclusion in \citet{neu12fsaur}.
}]
Judging only from the morphology of outbursts,
there does not seem to be an indication of truncation
of the disk.  Some outbursts of FS Aur are longer than
others, but superhumps have not yet been definitely
detected.  Both the IP status of this object and
the possible presence of a superoutburst or superhumps
need to be explored further.

\section{Summary}\label{sec:summary}

   We observed the 2014 superoutburst of the SU UMa-type
intermediate polar CC Scl.  We detected superhumps
with a mean period of 0.05998(2)~d during the superoutburst
plateau and during three nights after the fading.
During the post-superoutburst stage after three nights,
a stable period of 0.059523(6)~d was detected.  In addition to
these periods, we found that this object has shallow eclipses
and reached the identification of the orbital
period of 0.058567233(8)~d by using the available data
since 2011 and the CRTS data in quiescence.  We identified
the superhump period during the superoutburst plateau
to be stage B superhumps (according to the definition by
\cite{Pdot}) and post-superoutburst superhumps to be
stage C superhumps.  Such a late transition to stage C
superhumps has not been observed in other systems
and we consider that premature quenching of
the superoutburst may be responsible for this phenomenon.
By adopting the experimentally determined disk radii
in other SU UMa-type dwarf novae in the post-superoutburst
phase, we obtained a mass ratio of $q$=0.072(3) from
the dynamical precession rate of the accretion disk.
A modeling of the eclipse profile during outbursts
yielded an inclination of \timeform{80.6D}$\pm$\timeform{0.5D}.
The 2014 superoutburst was preceded by a precursor
outburst and the overall appearance of the outburst
was similar to a superoutburst in ordinary SU UMa-type
dwarf novae.  We discuss that the standard thermal-tidal
instability model can explain the outburst behavior in
this system and suggest that inner truncation of the disk
by magnetism of the white dwarf does not strongly affect
the behavior in the outer part of the disk.
Spin modulations were also recorded during outbursts,
and were enhanced by a factor of 15--30 compared to
the post-superoutburst state.  This can be naturally
explained by the increased accretion rate during outbursts.

\medskip

This work was supported by the Grant-in-Aid
``Initiative for High-Dimensional Data-Driven Science through Deepening
of Sparse Modeling'' from the Ministry of Education, Culture, Sports, 
Science and Technology (MEXT) of Japan.
We are grateful to the Catalina Real-time Transient Survey
team for making their photometric database available to the public.
We acknowledge with thanks the variable star
observations from the AAVSO International Database contributed by
observers worldwide and used in this research.
We thank an anonymous referee for improving the paper.
We thank K. Isogai and T. Ohshima for helping the compilation
of the observation.


\begin{thebibliography}{}

\bibitem[Andronov(1991)]{and91fsaur}
  Andronov, I.~L.\ 1991, IBVS, 3614, 1

\bibitem[{Araujo-Betancor} et~al.(2005)]{ara05v455and}
  {Araujo-Betancor}, S., {et~al.}\ 2005, A\&A, 430, 629

\bibitem[Augusteijn, Wisotzki(1997)]{aug97bwscl}
  Augusteijn, T., \& Wisotzki, L.\ 1997, A\&A, 324, L57

\bibitem[Baba et~al.(2000)]{bab00v1028cyg}
  Baba, H., Kato, T., Nogami, D., Hirata, R., Matsumoto, K., \& Sadakane, K.\
  2000, PASJ, 52, 429

\bibitem[Bianchini et~al.(1986)]{bia86gkper}
  Bianchini, A., Sabbadin, F., Favero, G.~C., \& Dalmeri, I.\ 1986, A\&A, 160,
  367

\bibitem[Chen et~al.(2001)]{che01ECCV}
  Chen, A., O'Donoghue, D., Stobie, R.~S., Kilkenny, D., \& Warner, B.\ 2001,
  MNRAS, 325, 89

\bibitem[{Cleveland}(1979)]{LOWESS}
  {Cleveland}, W.~S.\ 1979, J. Amer. Statist. Assoc., 74, 829

\bibitem[{de Martino} et~al.(2005)]{demar05htcam}
  {de Martino}, D., {et~al.}\ 2005, A\&A, 437, 935

\bibitem[{Drake} et~al.(2009)]{CRTS}
  {Drake}, A.~J., {et~al.}\ 2009, ApJ, 696, 870

\bibitem[{Evans}, {Hellier}(2005)]{eva05htcam}
  {Evans}, P.~A., \& {Hellier}, C.\ 2005, MNRAS, 359, 1531

\bibitem[Fernie(1989)]{fer89error}
  Fernie, J.~D.\ 1989, PASP, 101, 225

\bibitem[{Ge{\ss}ner}(1989)]{ges89fsaur}
  {Ge{\ss}ner}, H.\ 1989, Mitteil.\ Ver{\"{a}}nderl.\ Sterne, 11, 186

\bibitem[Hellier(2001)]{hel01book}
  Hellier, C.\ 2001, Cataclysmic Variable Stars: How and why they vary (Berlin:
  Springer)

\bibitem[{Hirose}, {Osaki}(1990)]{hir90SHexcess}
  {Hirose}, M., \& {Osaki}, Y.\ 1990, PASJ, 42, 135

\bibitem[Ishioka et~al.(2001)]{ish01j2315}
  Ishioka, R., Kato, T., Matsumoto, K., Uemura, M., Iwamatsu, H., \& Stubbings,
  R.\ 2001, IBVS, 5023

\bibitem[Ishioka et~al.(2002)]{ish02htcam}
  Ishioka, R., {et~al.}\ 2002, PASJ, 54, 581

\bibitem[{Kato} et~al.(2014a)]{Pdot6}
  {Kato}, T., {et~al.}\ 2014a, PASJ, in press (arXiv/1406.6428)

\bibitem[{Kato} et~al.(2013a)]{Pdot4}
  {Kato}, T., {et~al.}\ 2013a, PASJ, 65, 23

\bibitem[{Kato} et~al.(2014b)]{Pdot5}
  {Kato}, T., {et~al.}\ 2014b, PASJ, 66, 30

\bibitem[{Kato} et~al.(2009a)]{Pdot}
  {Kato}, T., {et~al.}\ 2009a, PASJ, 61, S395

\bibitem[{Kato}, {Maehara}(2013)]{kat13j1924}
  {Kato}, T., \& {Maehara}, H.\ 2013, PASJ, 65, 76

\bibitem[{Kato} et~al.(2010)]{Pdot2}
  {Kato}, T., {et~al.}\ 2010, PASJ, 62, 1525

\bibitem[{Kato} et~al.(2013b)]{kat13j1222}
  {Kato}, T., {Monard}, B., {Hambsch}, F.-J., {Kiyota}, S., \& {Maehara}, H.\
  2013b, PASJ, 65, L11

\bibitem[{Kato}, {Osaki}(2013)]{kat13qfromstageA}
  {Kato}, T., \& {Osaki}, Y.\ 2013, PASJ, 65, 115

\bibitem[{Kato} et~al.(2009b)]{kat09j0804}
  {Kato}, T., {et~al.}\ 2009b, PASJ, 61, 601

\bibitem[{Kato}, {Uemura}(2012)]{kat12perlasso}
  {Kato}, T., \& {Uemura}, M.\ 2012, PASJ, 64, 122

\bibitem[Kemp et~al.(2002)]{kem02htcam}
  Kemp, J., Patterson, J., Thorstensen, J.~R., Fried, R.~E., Skillman, D.~R.,
  \& Billings, G.\ 2002, PASP, 114, 623

\bibitem[{Mukadam} et~al.(2010)]{muk10v386ser}
  {Mukadam}, A.~S., {et~al.}\ 2010, ApJ, 714, 1702

\bibitem[{Neustroev} et~al.(2012)]{neu12fsaur}
  {Neustroev}, V., {et~al.}\ 2012, Mem.\ Soc.\ Astron.\ Ital., 83, 724

\bibitem[{Neustroev} et~al.(2013)]{neu13fsaur}
  {Neustroev}, V.~V., {Tovmassian}, G.~H., {Zharikov}, S.~V., \& {Sjoberg}, G.\
  2013, MNRAS, 432, 2596

\bibitem[Nogami et~al.(1998)]{nog98swuma}
  Nogami, D., Baba, H., Kato, T., \& Nov\'{a}k, R.\ 1998, PASJ, 50, 297

\bibitem[{Osaki}(1989)]{osa89suuma}
  {Osaki}, Y.\ 1989, PASJ, 41, 1005

\bibitem[{Osaki}(1996)]{osa96review}
  {Osaki}, Y.\ 1996, PASP, 108, 39

\bibitem[{Osaki}, {Kato}(2013a)]{osa13v1504cygKepler}
  {Osaki}, Y., \& {Kato}, T.\ 2013a, PASJ, 65, 50

\bibitem[{Osaki}, {Kato}(2013b)]{osa13v344lyrv1504cyg}
  {Osaki}, Y., \& {Kato}, T.\ 2013b, PASJ, 65, 95

\bibitem[Patterson(1994)]{pat94ipreview}
  Patterson, J.\ 1994, PASP, 106, 209

\bibitem[{Patterson} et~al.(1996)]{pat96alcom}
  {Patterson}, J., {Augusteijn}, T., {Harvey}, D.~A., {Skillman}, D.~R.,
  {Abbott}, T.~M.~C., \& {Thorstensen}, J.\ 1996, PASP, 108, 748

\bibitem[Patterson et~al.(1992)]{pat92dodra}
  Patterson, J., Schwartz, D.~A., Pye, J.~P., Blair, W.~P., Williams, G.~A., \&
  Caillault, J.-P.\ 1992, ApJ, 392, 233

\bibitem[Patterson, Szkody(1993)]{pat93dodraXray}
  Patterson, J., \& Szkody, P.\ 1993, PASP, 105, 1116

\bibitem[Schwope et~al.(2000)]{sch00RASSID}
  Schwope, A., {et~al.}\ 2000, Astron.\ Nachr., 321, 1

\bibitem[{Silvestri} et~al.(2012)]{sil12v455andGALEX}
  {Silvestri}, N.~M., {Szkody}, P., {Mukadam}, A.~S., {Hermes}, J.~J.,
  {Seibert}, M., {Schwartz}, R.~D., \& {Harpe}, E.~J.\ 2012, AJ, 144, 84

\bibitem[Skidmore et~al.(2000)]{ski00wzsge}
  Skidmore, W., Mason, E., Howell, S.~B., Ciardi, D.~R., Littlefair, S., \&
  Dhillon, V.~S.\ 2000, MNRAS, 318, 429

\bibitem[Stellingwerf(1978)]{PDM}
  Stellingwerf, R.~F.\ 1978, ApJ, 224, 953

\bibitem[{Tappert} et~al.(2004)]{tap04CTCV}
  {Tappert}, C., {Augusteijn}, T., \& {Maza}, J.\ 2004, MNRAS, 354, 321

\bibitem[Thorstensen et~al.(1996)]{tho96Porb}
  Thorstensen, J.~R., Patterson, J.~O., Shambrook, A., \& Thomas, G.\ 1996,
  PASP, 108, 73

\bibitem[{Tibshirani}(1996)]{lasso}
  {Tibshirani}, R.\ 1996, J. R. Statistical Soc. Ser. B, 58, 267

\bibitem[Tovmassian et~al.(1998)]{tov98htcam}
  Tovmassian, G.~H., {et~al.}\ 1998, A\&A, 335, 227

\bibitem[{Vogt}(1983)]{vog83lateSH}
  {Vogt}, N.\ 1983, A\&A, 118, 95

\bibitem[Warner(1995)]{war95book}
  Warner, B.\ 1995, Cataclysmic Variable Stars (Cambridge: Cambridge University
  Press)

\bibitem[{Warner}, {Pretorius}(2008)]{war08DNOQPO}
  {Warner}, B., \& {Pretorius}, M.~L.\ 2008, MNRAS, 383, 1469

\bibitem[Watson et~al.(1985)]{wat85gkperspin}
  Watson, M.~G., King, A.~R., \& Osborne, J.\ 1985, MNRAS, 212, 917

\bibitem[{Woudt} et~al.(2012)]{wou12ccscl}
  {Woudt}, P.~A., {et~al.}\ 2012, MNRAS, 427, 1004

\bibitem[{Zharikov} et~al.(2013)]{zha13ezlyn}
  {Zharikov}, S., {Tovmassian}, G., {Aviles}, A., {Michel}, R.,
  {Gonzalez-Buitrago}, D., \& {Garcia-Diaz}, M.~T.\ 2013, A\&A, 549, A77

\end{thebibliography}
\end{document}